\documentclass[3p,times,onecolumn]{elsarticle}

\usepackage{amssymb}

\usepackage[figuresright]{rotating}
\usepackage{subfigure}

\begin{document}

\begin{frontmatter}




\title{Characterization of low temperature metallic magnetic calorimeters having gold absorbers with implanted $^{163}$Ho ions}


\author[a]{L. Gastaldo\corref{cor1}}
\author[a]{P. Ranitzsch}
\author[a]{F. von Seggern}
\author[a]{J.-P. Porst}
\author[a]{S. Sch\"afer}
\author[a]{C. Pies}
\author[a]{S. Kempf}
\author[a]{T. Wolf}
\author[a]{A.~Fleischmann}
\author[a]{C. Enss}

\address[a]{Kirchhoff Institute for Physics, Heidelberg University, INF 227, 69120 Heidelberg, Germany}

\author[b]{A. Herlert\fnref{1}}
\author[b,c]{K. Johnston}

\address[b]{CERN, Physics Department, 1211 Geneva 23, Switzerland}
\address[c]{Technische Physik, Universit\"at des Saarlandes, 66041 Saarbr\"ucken, Germany}

\cortext[cor1]{Corresponding author \\
e-mail address: Loredana.Gastaldo@kip.uni-heidelberg.de}

\fntext[1]{Present address: FAIR GmbH, Planckstr. 1, 64291 Darmstadt, Germany}

\begin{abstract}
For the first time we have
investigated the behavior of fully micro-fabricated low temperature
metallic magnetic calorimeters (MMCs) after undergoing an ion-implantation process. This experiment had the aim to show the possibility to perform a high precision calorimetric measurement of the energy spectrum following the electron capture of $^{163}$Ho
using MMCs having the radioactive $^{163}$Ho ions implanted in the absorber. The
implantation of $^{163}$Ho ions was performed at ISOLDE-CERN. The
performance of a detector that underwent an ion-implantation process is compared to the one of a detector without implanted ions. The results show
that the implantation dose of ions used in this experiment does not compromise the
properties of the detector. In addition an optimized detector design for
future $^{163}$Ho experiments is presented.
\end{abstract}


\end{frontmatter}


\section{Introduction}
\label{intro}

The spectral resolving power $E/\Delta E_{\mathrm{FWHM}}$, defined as the ration between the energy absorbed in the detector and the energy resolution as the full width at half maximum (FWHM) of the gaussian detector response, of state of the art metallic magnetic calorimeters (MMCs) \citep{TOP2005}
for soft x-rays is above 2000. For completely micro-fabricated detectors, we recently achieved an energy resolution of
 $\Delta E_{\mathrm{FWHM}}\,=\, 2.0\,$eV at a photon energy of $E\,=\,6\,$keV and a signal rise
time of $90\,$ns \citep{PiesLTD14}. These properties make MMCs
very interesting detectors to study low energy processes like the
atomic de-excitation following electron capture (EC) \citep{nucc_2010}. Among these decays the high precision measurement of the calorimetric EC
spectrum of the isotope $^{163}$Ho is extremely interesting as it can be used to study the feasibility of future experiments investigating the electron
neutrino mass \citep{163Ho} \citep{Novikov-KlugeNNP} like MARE (Microcalorimeter Arrays for a Rhenium Experiment) \citep{MARE} and ECHO (Electron Capture Ho) experiment \citep{ECHO} .

To perform a calorimetric total activity measurement, the isotope
of interest needs to be surrounded by enough absorber material so
that the stopping power for the emitted particles is as close to 100$\%$ as possible.
In the experiment discussed here, this has been achieved by implanting $^{163}$Ho ions into a thin surface layer of a planar particle absorber made of gold
which afterwards is covered by a second
gold layer. The thickness of the two gold layers is defined to ensure a quantum efficiency close to 100$\%$.
The ion implantation was performed at ISOLDE-CERN
\citep{kugler_2000}. A mass selected beam, generated after surface
ionization of a Ta-W target which has been irradiated with protons, was
used. Together with the $^{163}$Ho other isobaric ions or ionized isobaric
molecules might get implanted at the same time.

The implantation process could potentially degrade the performance of the
detector.
Lattice defects are created by the high energy ions hitting
the absorber and could degrade its thermal properties. Moreover, the magnetic moments of the $^{163}$Ho ions as well as of the other contaminant ions that can be simultaneously implanted, might give rise to an additional heat capacity of the absorber which in turn would lead to a decrease of the expected temperature signal.

In the
following we compare the results obtained with a detector having an
ion-implanted absorber with results obtained with a detector taken
from the same fabrication run prior to the implantation.

\section{Metallic magnetic calorimeter}

\begin{figure}
  \includegraphics[angle=0, width=.40\textwidth]{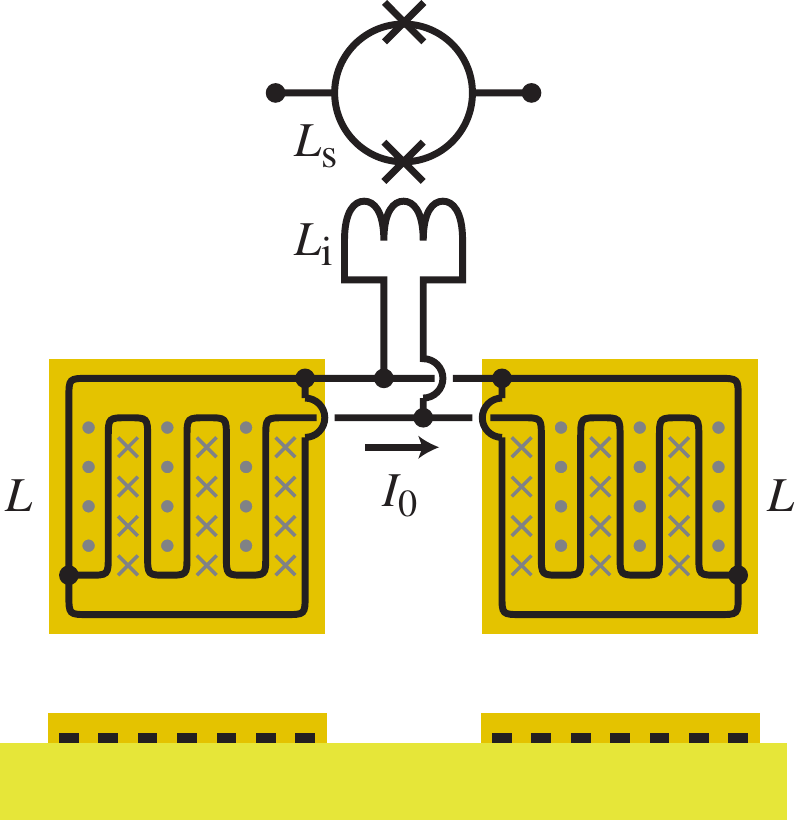}
  \caption{Schematic detector layout and SQUID coupling of the discussed micro-fabricated MMCs. The details are described in the text.\vspace{-1mm}}
  \label{DetectorScheme}
\end{figure}

Metallic magnetic calorimeters are energy dispersive
detectors operated at temperatures below $100\,$mK. They make use of the calorimetric
principle where the absorption of energy produces an increase
of the detector temperature proportional to the deposited energy $E$ and to the
inverse of the detector heat capacity $C_{\mathrm{tot}}$. A paramagnetic temperature
sensor, which resides in a small magnetic field, is tightly
connected to the particle absorber and weakly connected to a thermal bath. The change of temperature leads to a
change of magnetization of the sensor which can be read out as
a change of flux by a low-noise high-bandwidth dc-SQUID. The
sensor material, presently used for MMCs, is a dilute alloy of erbium in gold,
Au:Er. The concentration of erbium ions in the sensor can be chosen to
optimize the detector performance and usually has a value between 200
ppm and 800 ppm.

The detector geometry used in the experiment discussed here is based on a planar superconducting meander-shaped pick-up coil \citep{FleischLTD13}. A schematic of this design is shown
in Fig.\ref{DetectorScheme}. Two meander-shaped pick-up coils, each of inductance $L$, are connected in parallel and both are in parallel to the input coil of a current sensor dc-SQUID having an inductance $L_{\mathrm{i}}$. Regarding the magnetic flux coupled to the SQUID, when the two pick-up coils are exposed to an external magnetic field , they behave like a first order gradiometer. A planar Au:Er sensor is located on
top of both superconducting meander-shaped pick-up coils (according to experiment requirements one of the sensor can also be omitted).

A persistent current $I_{\mathrm{0}}$ can be prepared in the superconducting loop formed by the two meander-shaped pick-up coils. This current provides the magnetic field used to polarize the magnetic moments in the sensor. The change of flux in one of the two meander-shaped pick-up coils, following the interaction of a particle in the corresponding pixel, generates screening currents in all the three inductances, i. e. the two meander shaped pick-up coils and $L_{\mathrm{i}}$. The screening current flowing in the input coil is then generating a flux in the dc-SQUID of inductance $L_{\mathrm{S}}$. The flux signal is transduced in a voltage signal, which, after being filtered and amplified, will be acquired and analyzed.

\section{Detector design and fabrication}
\label{MMCs}

\begin{figure*}
  \includegraphics[angle=0, width=.99\textwidth]{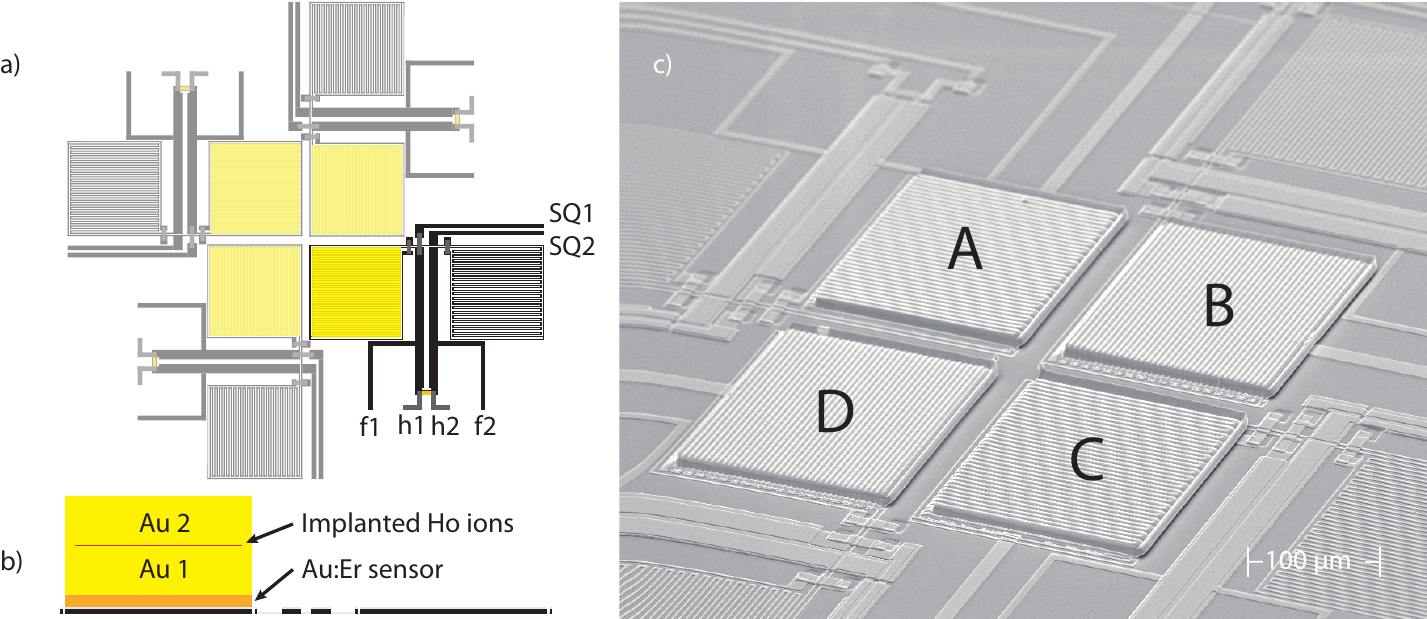}
  \caption{Schematic and SEM picture of a MMC detector chip for $^{163}$Ho implantation. a) Layout of the four double-meander pick up coils. b) Schematic of the absorber cross-section. c) SEM picture of the central part of the fully micro-fabricated MMC with the four pixels (A, B, C and D)  prepared to be irradiated \vspace{-3mm}}
  \label{4pixels}
\end{figure*}

Fig. \ref{4pixels} a) shows the layout of the detector chip developed for the first ion-implantation test. Each chip is equipped with four detectors. The detectors are based on the niobium double-meander pick-up coil geometry. The leads used to inject the persistent current $I_{\mathrm{0}}$ into the meander-shaped pick-up coils are labelled with f1 and f2 respectively. The persistent current in the meander is injected by means of an on-chip persistent current switch which is formed by an U-shaped extension of the meander circuit. Part of this extension, far away from the meander structure, can be heated above the critical temperature of niobium, $T_{\mathrm{c}}\, \approx \,9\,$K, by means of a resistive heater. The lines used to run a current through the resistive heater of the on-chip persistent current switch are labelled with h1 and h2, respectively. Finally, the leads to connect the double-meander structure to the input coil of the dc-SQUID are labelled SQ1 and SQ2, respectively.

In order to be able to measure the temperature and magnetic field dependence of the magnetization of the paramagnetic sensor material and to compare it to the expected behavior, only one of the two meander-shaped pick-up coils of each read-out channel was covered with a Au:Er sensor.

The four single pixels are positioned in a way to be close to each other so that the implantation could
be performed into all of them at the same time. The erbium concentration and geometry of the sensor are optimized to reach high energy resolution when combined with a gold  absorber having the dimension of $190\times 190\times 10\, \mu \mathrm{m} ^3$. The absorber dimensions are suitable for having a quantum efficiency close to $100\, \%$ for the radiation emitted during the EC of $^{163}$Ho.

Fig. \ref{4pixels} b) shows the cross-section of one of the detectors. In particular, the position of the implanted ions is indicated as a solid line between the two parts of the  gold absorber, Au1 and Au2. The ions are implanted within an area of $160\times160 \,\mu$m$^2$ after the production of the first half of the absorber, Au1, a $5\, \mu \mathrm{m}$ thick gold film.
The second part of the absorber, Au2,
a $5 \,\mu \mathrm{m}$ thick gold film, is structured in a succeeding step over the full absorber area,
$190 \times 190\, \mu \mathrm{m}^2$. The foreseen energy resolution achievable by this detector, considering the bulk thermal properties of gold and Au:Er
 when coupled to a suitable dc-SQUID with matched input inductance,
and with no unpredictable noise source or temperature instability, is about $4\,$eV FWHM.

Fig \ref{4pixels} c) shows a scanning
electron microscope (SEM) picture of the central part of the
detector chip. The first gold layer of the absorber was processed to be ready for ion implantation.

\subsection {Micro-fabrication}
\label{microf} The detector chips are micro-fabricated on a $430\, \mu$m thick thermally oxidized 2-inch
silicon wafer. The meander-shaped pick-up coils and part of the leads are structured using a wet-etching process starting from a continuous $400\,$nm thick niobium film. Each meander-shaped pick-up coil covers an area of
$190\times190\,\mu\mathrm{m}^2$. The pitch is $5\,\mu$m with a meander-line
width of $2.5\,\mu$m. The critical current of the superconducting
meander structure was measured to be $130\,\mathrm{mA}$ for chips prior to the implantation. The  surface of the niobium structures gets electrically isolated by an anodization process and an additional $200\, \mathrm{nm}$ thick SiO$_\mathrm{x}$ layer. Parts of the niobium film, like bonding pads and vias, are left free from isolation. The electrical circuits get closed through vias and lines in a second $600\,$nm thick niobium layer.
The resistive heater of the persistent current switch is
made of sputtered $75\,$nm thick AuPd with a $5\,$nm thick titanium sticking
layer. The resistance of the heater was measured to be about $30\,\Omega$. Reliable switching and current injection is achieved
with heater currents of about $2.5\,$ mA. The Au:Er sensor is
sputter-deposited onto the meander structures. A thin sticking
layer of less than $5 \,$nm of 
niobium is used. To obtain the
desired erbium concentration, a co-sputtering process from a pure gold
target (5N) and a Au:Er target with $840\,$ppm erbium (which was produced
from 5N purity gold and isotopically enriched $^{168}$Er) is
applied. The concentration of erbium in the sensor of the
detectors discussed here is 255 ppm. This value was obtained by measuring the
magnetization of the Au:Er foil, stripped from the photo-resist surface during the lift-off process, in a commercial
magnetometer (MPMS 5X, Quantum Design \citep{magnetometer}) within the temperature range from 300 K to
$2\,$K. The height of the sensor is $1.3 \,\mu$m.
In general the absorber
can have mechanical and electrical contact to the sensor either on the complete surface of the sensor or only on a smaller fraction of it through micro-fabricated stems. The absorber
could have the same area as the sensor or it could be overhanging.
The detectors developed to be irradiated were designed to have the absorbers as large as the sensors and with a connection to each other on the complete area.
This design allows for the maximal mechanical stability in the finishing steps of the absorber
preparation, following the implantation process.
After the fabrication of the Au:Er sensor a $5\,\mu$m
thick gold absorber is produced by electroplating gold into a
photo-resist mold using the commercial electrolyte Techni-Gold 25E RTU \citep{gold_electrolite}. Some further details about the micro-fabrication steps can be found
in \citep{FleischLTD13}. After the production of the $5 \,\mu$m thick gold
absorbers, the wafer is diced into detector chips. The size of each detector chip is $5\times5 \,$mm$^2$. A
chip prepared with the described procedures has been tested. One
of the four detectors was characterized at low temperature by
analyzing the thermal response upon the absorption of x-rays from
an external $^{55}$Fe calibration source and the results will be discussed in Section 5.

\subsection{Ion implantation and absorber completion}
\label{ion}

A single chip, equipped with four detector pixels, was
glued into a $5\times 5\,$mm$^2$ hole of a glass mask in a way to have a suitable large surface at
the level of the chip surface itself for better manipulation and feasibility of the further
micro-fabrication steps. A photo-resist layer was structured to
protect all parts of the chip besides four squares, each with a surface
of $160\times160\,\mu$m$^2$, each of them positioned in the center of an
absorber. These squares define the region where the implantation
will take place. The implantation surface on each absorber was kept
smaller than the area of the absorber itself in order to ensure,
once the absorber is completed,
that no radiation leak along the side walls occurs.
After this
step the chip was sent to the ISOLDE radioactive ion-beam facility at CERN for the
ion implantation.

The radioactive isotopes were produced using the ISOL technique
\citep{Ravn_98}. A $1.4 \,$GeV pulsed proton beam was used to irradiate
a bulk metallic Ta-W target kept at a temperature of about $1800-1900^{\mathrm{o}}$C. After surface ionization, an ion beam containing different isotopes and
molecules is extracted at 30 kV and accelerated to the mass separator. The
detector chip was mounted at the collection chamber of the low mass beam-line connected to the GPS (General Purpose Separator) mass separator.
The mass-resolving power $R=m/\Delta m$ of the GPS is about 1000.
For the implantation the mass $m\,=\,163\,$u was selected. The resolving power is good enough to keep contamination of ions of different mass on a negligible level, but contamination of the
beam can come from isobaric ions and ionized isobaric molecules.

The detector chip
was irradiated for 40 minutes with a current of $75\,$pA corresponding to a number of monovalent ions of about $10^{12}$ of which about $10^{11}$ were implanted into the four absorbers. The
diameter of the beam was about $2\,$ mm. After the implantation the
sample was stored for a few months to bring the residual radioactivity due to short living isotopes to a negligible level. After this storing time a contamination test was performed. The results of this test will be discussed in section 4.

The absorber was completed about three months after the
implantation. Two different gold depositions were performed. A first
very thin gold layer, about $100\,$nm thick, was thermally evaporated at the
ISOLDE facility. Before this evaporation, no additional
micro-structuring steps were done on the chip.
After this step, still at ISOLDE, the photo-resist and the residual gold layer were removed leaving the detector chip free of any radioactive parts besides the absorbers. The thin gold protecting layer and the removal of photo-resist were done at CERN in order
not to deal with possible contaminating material while processing
the detector chip in not suitably equipped laboratories in the Kirchhoff-Institute for Physics at Heidelberg University.

The so-prepared detector chip was sent back to the Kirchhoff-Institute for Physics for the finishing steps. After the spin-coating and the photo-lithography process of a photo-resist layer, a
 $5\,\mu$m thick Au film was sputter-deposited all over the four
absorber areas. A final lift-off process made the detector chip ready to be measured in a cryostat.

\subsection{Detector set-up and read-out} \label{read}
In the two experiments described here, only one pixel per detector chip has been tested.
The detector chip and the front-end SQUID are
glued to a brass holder. The SQUID input coil is connected with
wedge-bonded aluminum wires in parallel to the meander-shaped pick-up coils  as schematically
shown in Fig. \ref{DetectorScheme}. Superconducting leads present on the brass holder are wedge-bonded to the heater and persistent current leads on the detector chip. Additional leads are used for the operation of the dc-SQUID. The brass holder is
inserted into a superconducting lead shield to screen time-dependent external magnetic fields. The shield has a small hole positioned above a collimator to allow the x-rays from the external $^{55}$Fe
calibration source to reach the absorber. The screened
detector holder is then mounted at the cold stage of a
commercial adiabatic demagnetization refrigerator (ADR)
 which can reach a base temperature of about $23\,$ mK. The signal of the front-end SQUID is passed for a
first amplification to a second SQUID. More details of the 2-stage SQUID read-out scheme can be found in \citep{FleischLTD13}. This scheme ensures a
reduced apparent flux noise of the read-out chain referred to the front-end SQUID and a lower power dissipation in
the detector SQUID. The SQUID used as amplifier was mounted at the
experimental platform of the ADR. The signal was then
amplified by a room temperature SQUID-electronics,
XXF-1, Magnicom GmbH, Hamburg, Germany \citep{SQUID_PTB}. After being filtered, further amplified and digitized with a 12-bit ADC card, the signal is
acquired and a preliminary on-line analysis is performed using a software
developed in our group.
The two-stage SQUID read-out used in the discussed experiments was
the same for both detectors. The front-end SQUID was of the type
C4X1 \citep{SQUID_PTB} with a nominal input inductance of $1.8\,$nH. The
second stage SQUID was a sixteen SQUIDs series array of the type C5X16
\citep{SQUID_PTB}.

\begin{figure}[t!]
  \includegraphics[angle=0, width=.44\textwidth]{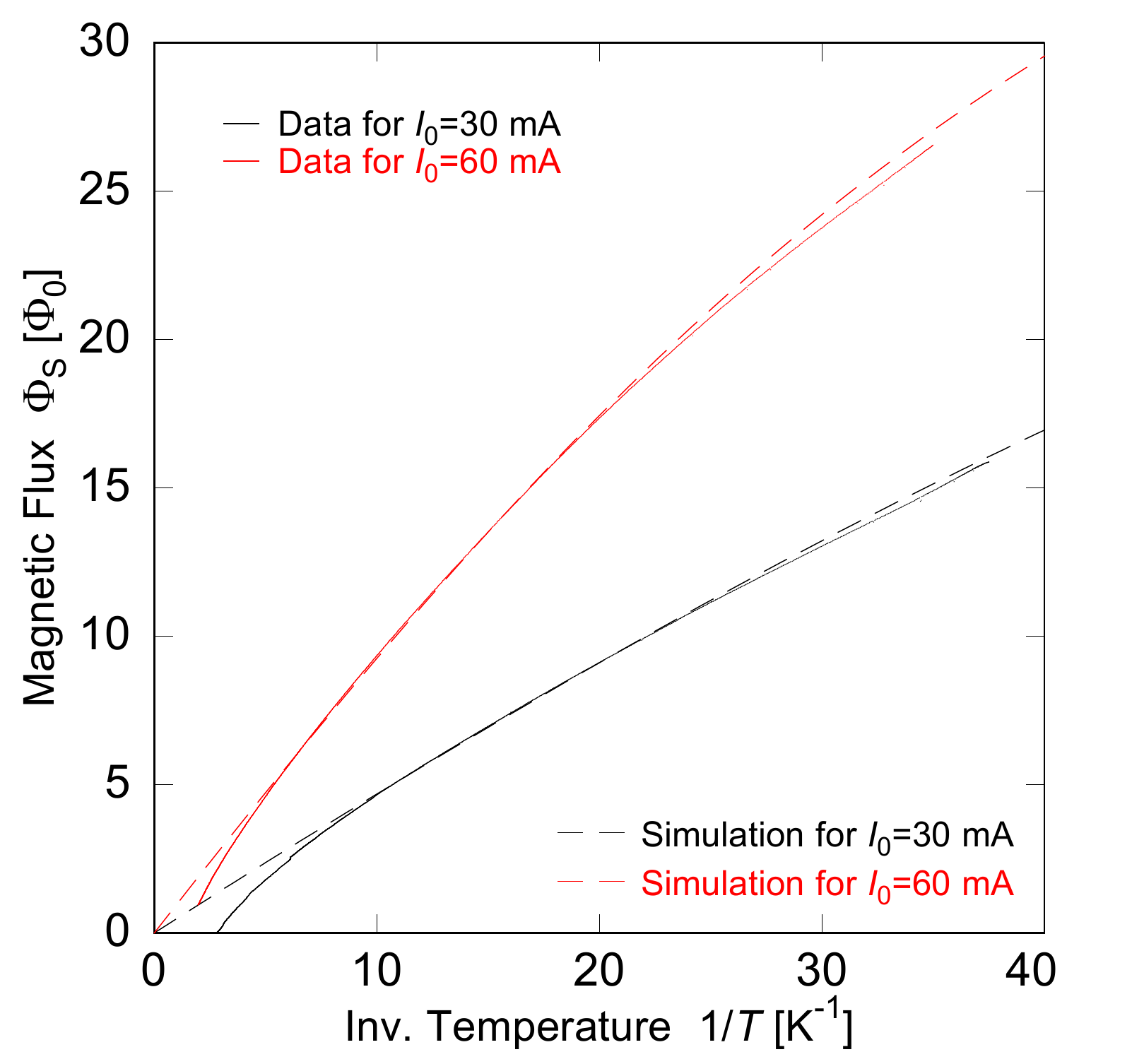}
  \caption{Measured dc-magnetization of the sputtered Au:Er$_{255\mathrm{ppm}}$ sensor for the  detector with non implanted absorber in units of magnetic flux in the front-end SQUID as a function of inverse temperature for two different persistent currents $I_{\mathrm{0}}$ in the meander-shaped pick-up coil. \vspace{-1mm}}
  \label{Magnetization Falk}
\end{figure}
\begin{figure}[t!]
  \includegraphics[angle=0, width=.44\textwidth]{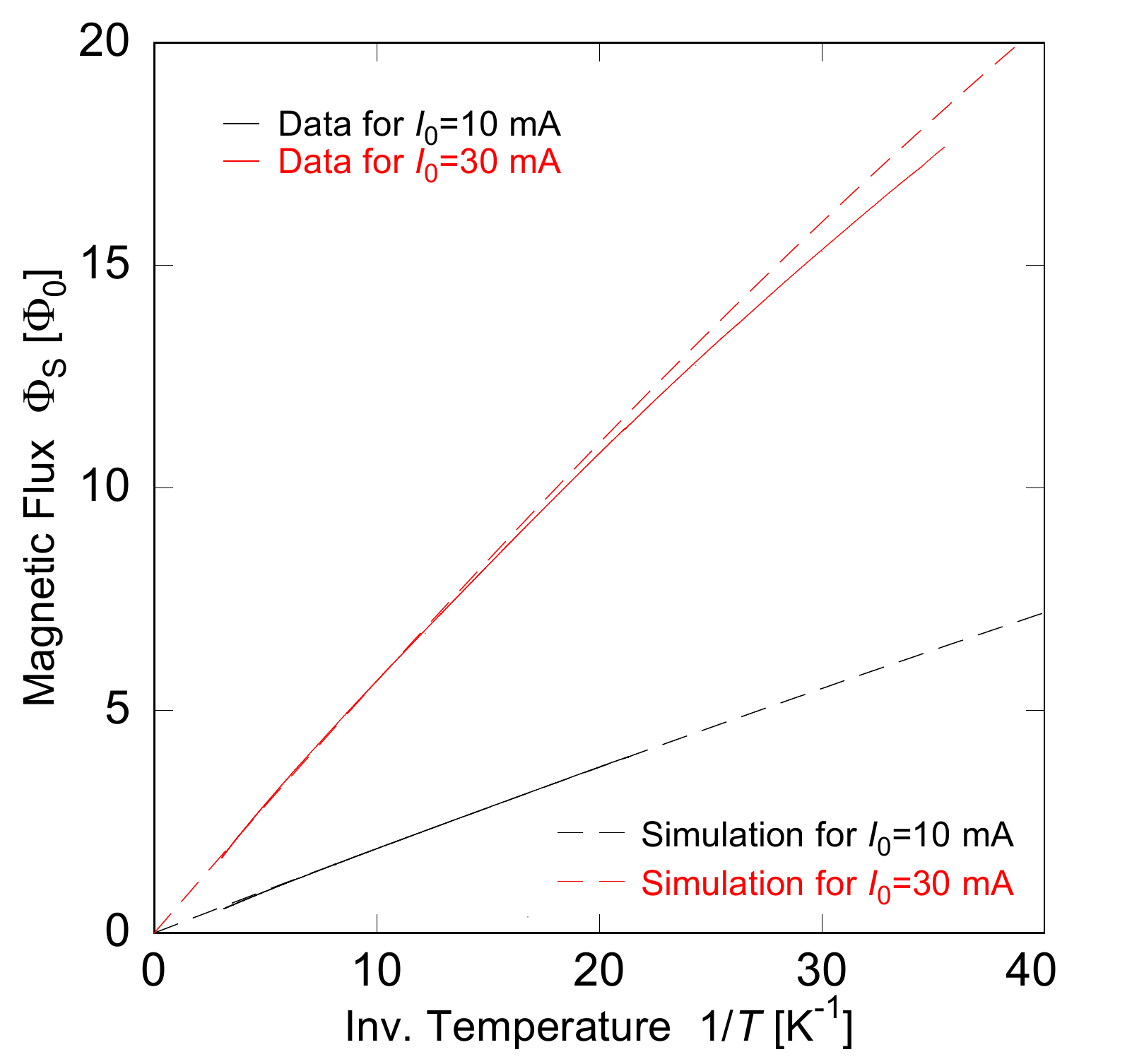}
  \caption{Measured dc-magnetization of the sputtered Au:Er$_{255\mathrm{ppm}}$ sensor for the  detector with implanted absorber in units of magnetic flux in the front-end SQUID as a function of inverse temperature for two different persistent currents $I_{\mathrm{0}}$ in the meander-shaped pick-up coil. \vspace{-1mm}}
  \label{Magnetization Philipp}
\end{figure}

\section{Contamination test}
\label{cont} The purpose of this experiment was to test the performance of a
micro-fabricated MMC detector after undergoing an ion-implantation process and to
provide a preliminary measurement of the $^{163}$Ho calorimetric EC
spectrum, therefore no dedicated study was done for the
preparation of high purity $^{163}$Ho beam. A contamination
measurement was performed a few months after the implantation took place.
All the samples that have been implanted in the same run (the detector chip and a gold foil) were exposed to a gamma detector (Ortec Well Ge detector) in
order to measure the emission of gamma rays.

This measurement
showed that the major contamination was from $^{147}$Gd with a
half-life of 38.06 h, $^{147}$Eu with a half-life of 24.1 d and
$^{144}$Pm with half-life of 363 d. The isotope $^{147}$Gd and part of the present ions of the isotope
$^{147}$Eu were implanted as ionized GdO$^+$ and EuO$^+$. The isotope
$^{147}$Gd gives rise to a decay chain $^{147}$Gd $\rightarrow$
$^{147}$Eu $\rightarrow$ $^{147}$Sm. $^{147}$Sm decays with the
emission of alpha particles with a half-life of $1.06\, 10^{11}\,$y.
Therefore the contribution of this isotope to the background spectrum of the discussed experiment can be considered as negligible. The isotope
$^{144}$Pm was also implanted as ionized molecule of mass $163\,$u, as PmF$^+$.  This isotope decays through EC or positron emission processes to
$^{144}$Nd, an alpha emitter with a half-life of $2.29\, 10^{15}\,$y.
The contribution of this last isotope to the background spectrum can here be considered
negligible. According to the half-life values, the main
contribution to the background of the discussed experiment, a few months after the
implantation, is given by $^{144}$Pm.

In future experiments for the investigation of the neutrino mass, the presence of radioactive contaminants in the $^{163}$Ho source needs to be eliminated. Because of that a detailed study on the preparation of a high pure $^{163}$Ho source already started. Different methods will be investigated: irradiation with charged particles of suitable targets at accelerator facilities \citep{Susanta} and neutron irradiation of an erbium target enriched with $^{162}$Er \citep{galeazzi}.

\section{Detector characterization}
The detector with ion-implanted $10\,\mu$m thick gold absorber as well the one with non-implanted
$5\,\mu$m thick gold absorber were tested by analyzing the magnetization of the Au:Er sensor and the response upon the absorption of x-rays from an external $^{55}$Fe calibration source. In the case of the detector with implanted ions, also the response to the de-excitation radiation generated after the EC process in $^{163}$Ho has been used for the analysis.

\subsection{Magnetization}
The low temperature magnetic properties of the Au:Er sensor were
tested by measuring the magnetization of the sensor as a function
of temperature for different field generating persistent currents in the meander-shaped pick-up coil.
Fig. \ref{Magnetization Falk} shows the magnetic flux in the primary SQUID caused by the
magnetization of the Au:Er sensor as a function of inverse bath temperature for the detector having the non-implanted absorber. The data for two different persistent currents in the meander-shaped pick-up coils (continuous lines) are compared to the theoretical expectation calculated numerically (dashed lines)as discussed in \citep{TOP2005}. For medium high temperatures, i.e. between $50\, $mK and $150\, $mK, the agreement between the measured and calculated data is very good. The small deviation at lower temperatures is most likely explained by thermal decoupling of the detector chip from the bath due to power input from the front-end SQUID. At temperatures above $150\, $mK the expected Curie-like behavior, $M\propto 1/T$, is not observed due to an insufficient shielding, which is unable to completely shield the stray field of the ADR  magnet.
Fig. \ref{Magnetization Philipp} shows the corresponding data of the detector with ion-implanted absorber.
For these data sets no deviation from the expected behavior can be seen at high
temperature, due to an improved shielding. At low temperatures a small thermal decoupling of the detector chip from the thermal bath is again observed. Comparing the results obtained with the two
detectors we conclude that the implantation process, as
expected, does not compromise the magnetic properties of the
sensor. In fact the implanted ions are at a relatively large distance from to the meander-shaped pick-up coil, about $6.3 \, \mu$m, and therefore the magnetic moments of those ions have a negligible coupling to the coil.
\begin{figure} [t]
  \includegraphics[angle=0, width=.44\textwidth]{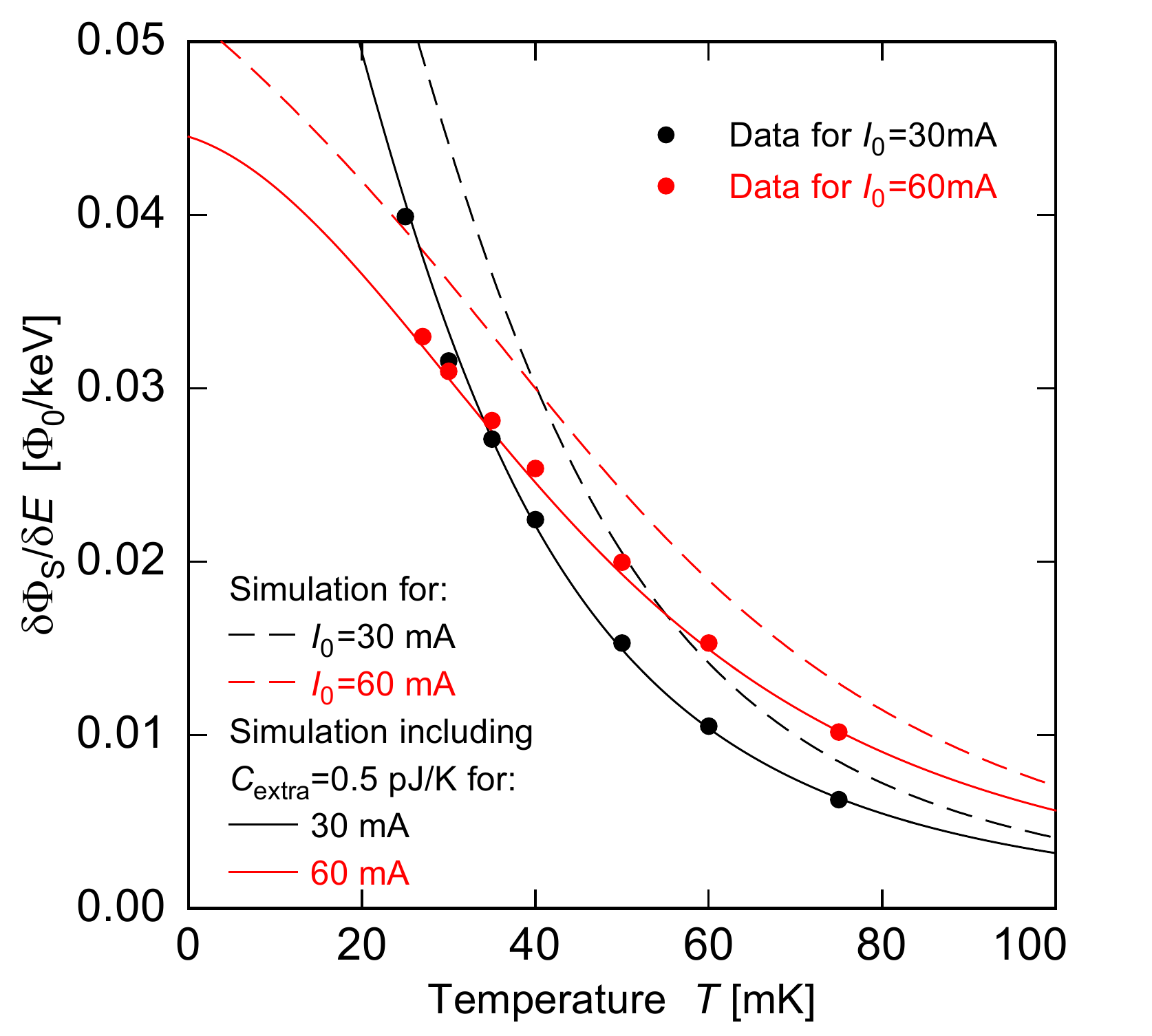}
  \caption{Measured pulse height (dots) for the non-irradiated detector compared to the expected values calculated using bulk material parameters (dashed lines) and to a theoretical prediction with included extra heat capacity for sputtered gold (solid lines)\vspace{-1mm}}
  \label{Pulse height Falk}
\end{figure}
\begin{figure} [t]
  \includegraphics[angle=0, width=.44\textwidth]{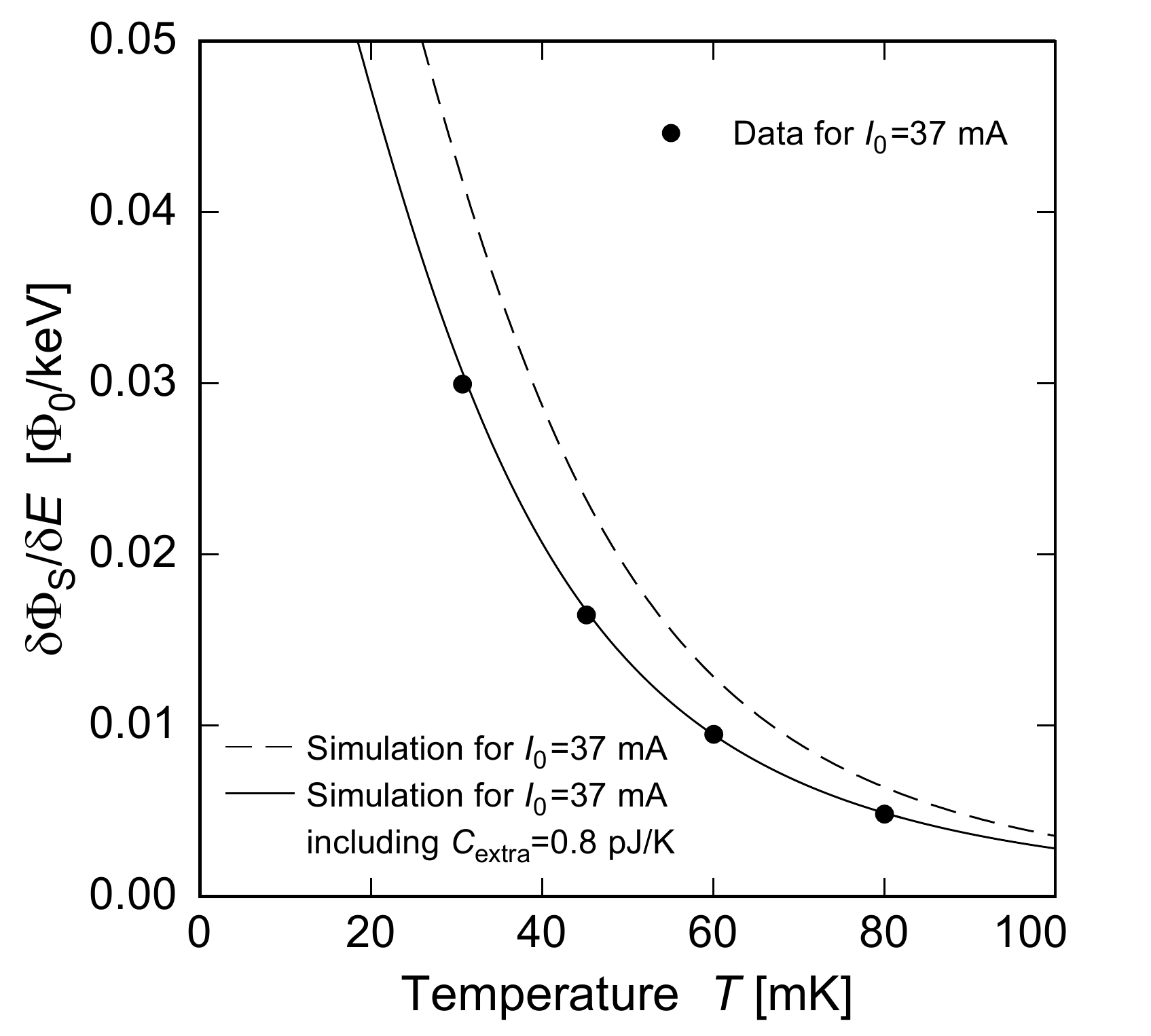}
  \caption{Measured pulse height (dots) for the irradiated detector compared to the expected values calculated using bulk material parameters (dashed line) and to a theoretical prediction with included extra heat capacity for sputtered gold (solid line)\vspace{-1mm}}
  \label{Pulse height Philipp}
\end{figure}
\subsection{Signal shape and size}

The implantation of heavy ions in a gold film produces lattice
defects which could change the thermal properties of the absorber,
in particular they could be responsible for an additional contribution in
the heat capacity. Moreover, the ions with large electronic magnetic moments and/or nuclear moments present in the implanted beam could add another contribution to the heat capacity due to interacting localized electronic magnetic moments and hyperfine split nuclear levels.
To test these effects, we compared the response upon the deposition of energy in the absorber of the detector with non-implanted absorber to the response of the detector with implanted absorber. We analyzed the shape of
the thermal pulses as a function of temperature and magnetic field.
The measured pulse heights, which are
proportional to the inverse of the total heat capacity of the
detector, are compared to the calculated values. The numerical
calculation includes, in case of sputtered gold and Au:Er films, an additional contribution to the specific heat to be added to the one of the corresponding bulk material. This effect has been discussed in \citep{FleischLTD13}. The additional specific heat per
volume seems to vary somewhat with details of the sputtering conditions and takes values between 2 and 10  J/K/m$^3$ where the largest values are found for sputtered Au:Er films and the smallest are found for sputtered Au films. Moreover, this additional contribution to the specific heat was found to be fairly temperature and magnetic field independent in the investigated range of the parameters.

Fig. \ref{Pulse height Falk} shows the pulse height per keV of
absorbed energy in the non-irradiated detector as function of
temperature for two different persistent currents in the meander-shaped pick-up coil.
The experimental data, indicated in the figure as dots, follow perfectly the expected behavior indicated by the solid lines. The theoretical curves have been calculated using an additional contribution to the heat capacity for the Au:Er film of $C_{\mathrm{extra}}\,=\,0.5\,$pJ/K, which corresponds to an additional specific heat of 10 J/K/m$^3$. No additional contribution of the heat capacity was used for the gold absorber since only the first half of the absorber exists in this device and this was produced by electro-plating technique. The dashed lines indicate the signal size that could be reached in case the heat capacity of the Au:Er film would be described using only the specific heat of bulk alloy.

Fig. \ref{Pulse height Philipp} shows the measured pulse height for
the detector with implanted absorber per keV of absorbed energy
versus temperature for a persistent current of $I_{\mathrm{0}}\,=\,37\,$mA in the meander-shaped pick-up coils. Also in this case the experimental data, indicated as dots, follow perfectly our expectations, indicated as solid line. The theoretical expectation were calculated including the extra heat capacity contribution of $0.5\,$pJ/K of the Au:Er film and $0.3\,$pJ/K for the second half of the gold absorber of since it was fabricated by sputter-deposition. The dashed line indicates the achievable pulse height in the absence of the additional contribution to the heat capacity of the sputtered gold and Au:Er films.

The thermal response of the detector with implanted absorber does not show to
be noticeably degraded by defects created during the implantation or by additional contributions to the heat capacity due to magnetic impurities.
Due to the
fact that the beam was not only containing $^{163}$Ho but other
isobars (atomic and ionized molecules), the contribution to the heat
capacity related to the magnetic moment of a pure $^{163}$Ho implantation could not be tested. On the other hand, we can argue that the largest part of the contaminants also carries magnetic moments. Therefore we can set a very important limit for the future detector design: the lattice defects and the presence of magnetic moments due to the implantation of about $10^{11}$ atoms in the detector absorber do not noticeably compromise the performance of the detector. We can conclude that activities of few Bq (about $2\, 10^{11}$ $^{163}$Ho ions correspond to 1 Bq) are allowed for the final design MMC for the high energy resolution calorimetric measurement of the EC spectrum of $^{163}$Ho. Dedicated experiments to find the maximum allowed activity per pixel are planned. These experiments will investigate the thermo-dynamical properties of dilute alloys having different holmium concentrations.

A second aspect that was tested is the pulse rise-time. Comparing the first $100\,\mu$s of a thermal pulse deriving from the detector with non-implanted absorber with a pulse from the detector with ion-implanted absorber, we could not detected any noticeable difference, in both cases the rise-time was about $0.1\,\mu$s.

\subsection{Energy resolution}

In order to characterize the non-irradiated detector, an energy
spectrum of x-rays emitted by an external $^{55}$Fe
calibration source was acquired. The detector was operated at a
temperature of $30\, \mathrm{mK}$ with a field generating persistent current
of $30\, \mathrm{mA}$.

\begin{figure}
  \includegraphics[angle=0, width=.44\textwidth]{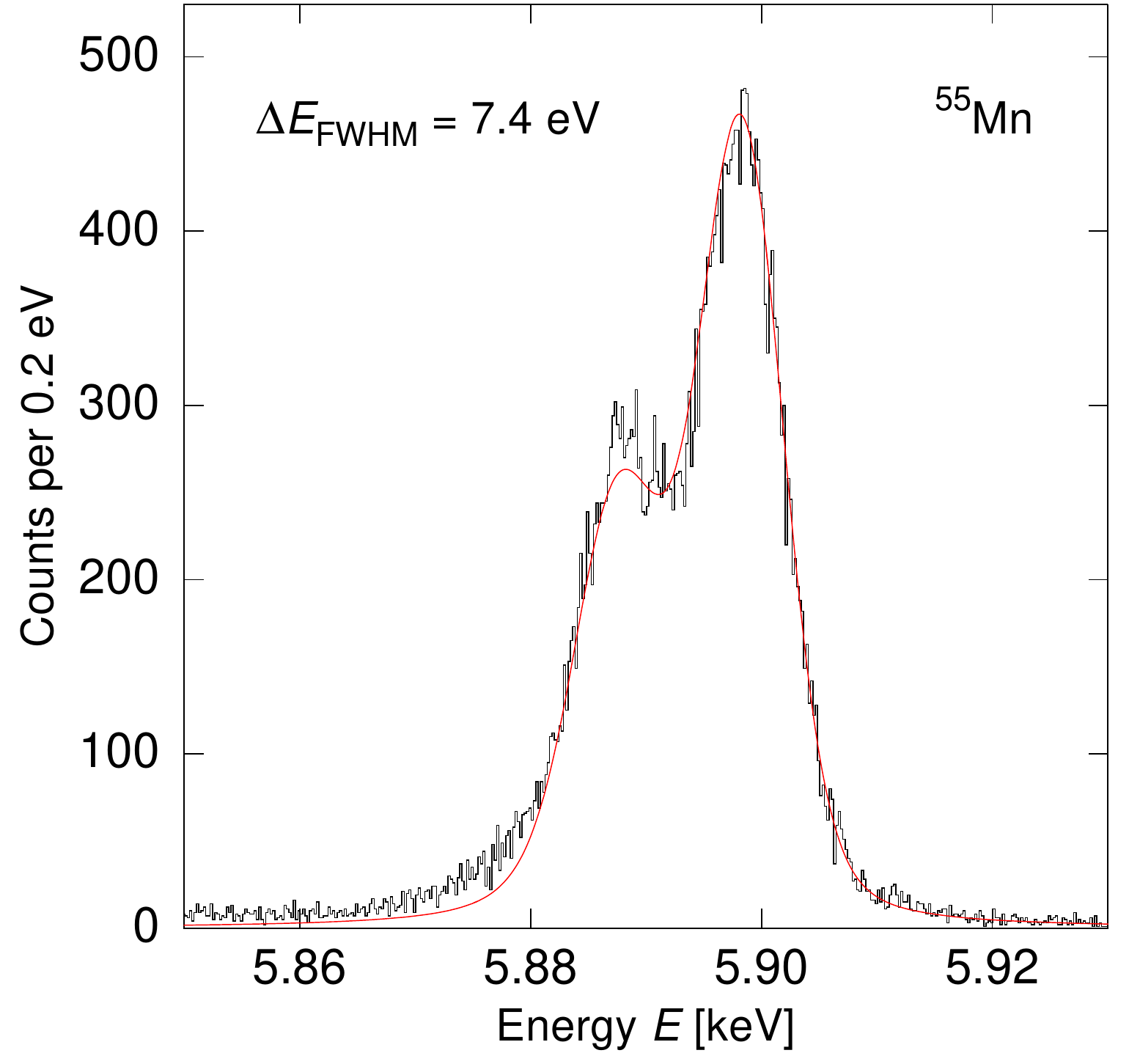}
  \caption{Spectrum of the $K_{\mathrm{\alpha}}$-line of Mn acquired with the MMC with non-implanted absorber. The fit function, obtained by convolving the natural line-shape described by the parameters given in \citep{Hoe97} with a gaussian having a width of 7.4 eV FWHM, is superimposed to the experimental data (red line).\vspace{-1mm}}
  \label{Mnka Falk}
\end{figure}

\begin{figure} [t]
  \includegraphics[angle=0, width=.44\textwidth]{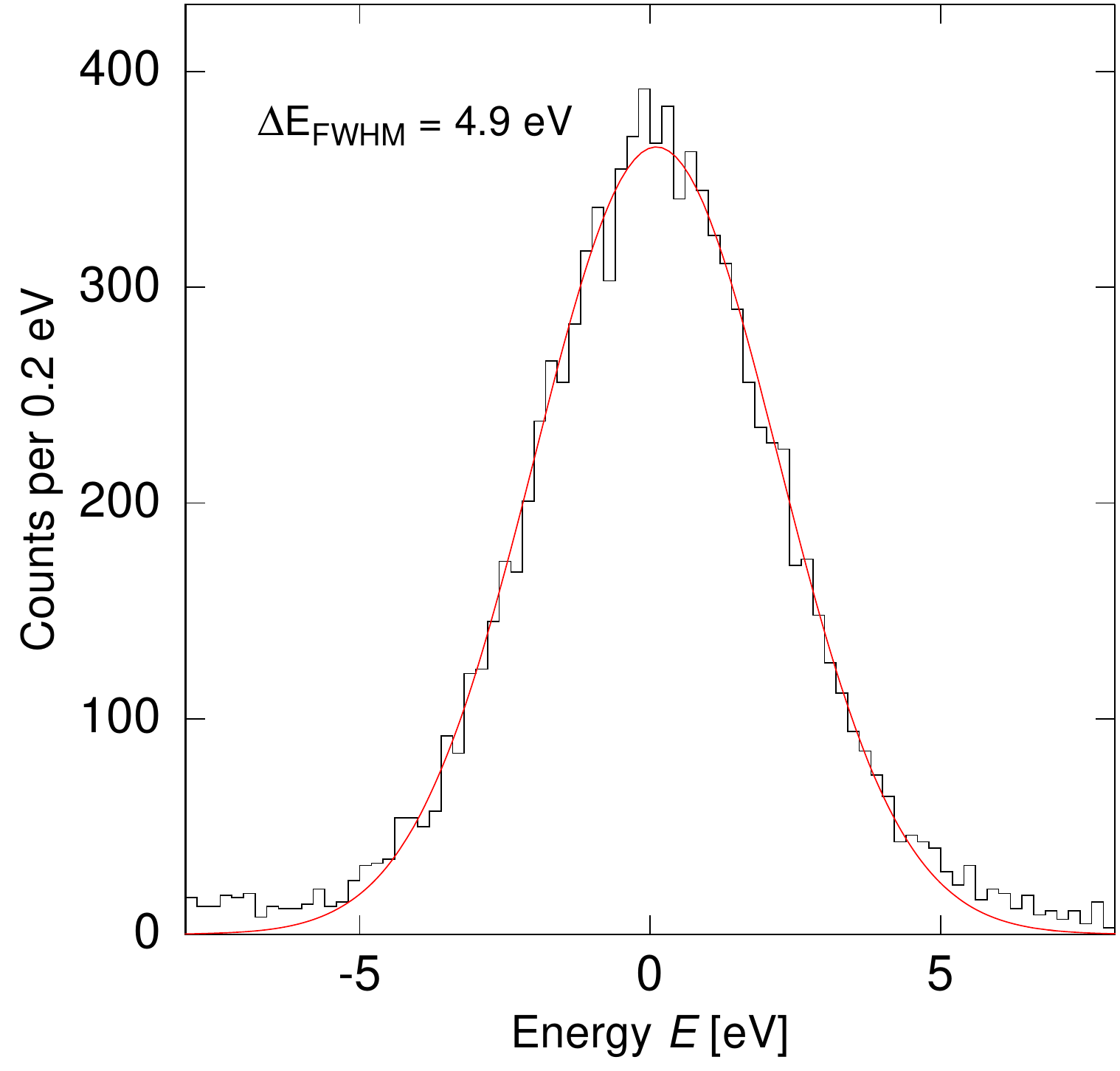}
  \caption{Baseline spectrum acquired with the MMC with non-implanted absorber. A Gaussian fit with a width of 4.9 eV (FWHM) is superimposed to the experimental data (red line).\vspace{-1mm}}
  \label{Ho baseline falk}
\end{figure}

Fig. \ref{Mnka Falk} shows the acquired spectrum of the $K_{\mathrm{\alpha}}$-line of manganese. A gaussian distribution with variable width, which represents the detector response, convolved with the natural line shape of the $K_{\mathrm{\alpha}}$-line of manganese \citep{Hoe97} was used to fit the experimental data. The energy resolution extracted from the fit shown in Fig. \ref{Mnka Falk} is $\Delta E_{\mathrm{FWHM}}\, = \,7.4 \,$eV. Fig. \ref{Ho baseline falk} shows a corresponding energy spectrum around $0 \,$eV (baseline spectrum) as derived from applying the same data reduction algorithm to un-triggered detector signal traces. A gaussian fit was performed on this data set. The width of the fit represents the intrinsic energy resolution of the detector of $\Delta E_{\mathrm{FWHM}}\,=\,4.9\,$eV. The difference in energy resolution between
the baseline spectrum and the measured $K_{\mathrm{\alpha}}$-line of manganese can be partially explained by  temperature variations of the experimental holder and by a partial loss
of the energy deposited by some of the incoming x-rays in the form of high energy
phonons which are transmitted directly to the substrate without increasing the temperature of the sensor. This explanation is
supported by the existence of an additional shoulder at the low energy
side of the $K_{\mathrm{\alpha}}$-line of manganese shown in the spectrum of Fig. \ref{Mnka Falk}. The loss of energy through high energy phonons can be
reduced by reducing the contact area between the sensor and the
absorber \citep{FleischLTD13}. Measurements using a detector designed for high resolution x-ray spectroscopy with a reduced contact
area, achieved by introducing stems between absorber and sensor, have shown a noticeably improved symmetry of the measured lines \citep{FleischLTD13}.

\begin{figure}[t]
  \includegraphics[angle=0, width=.44\textwidth]{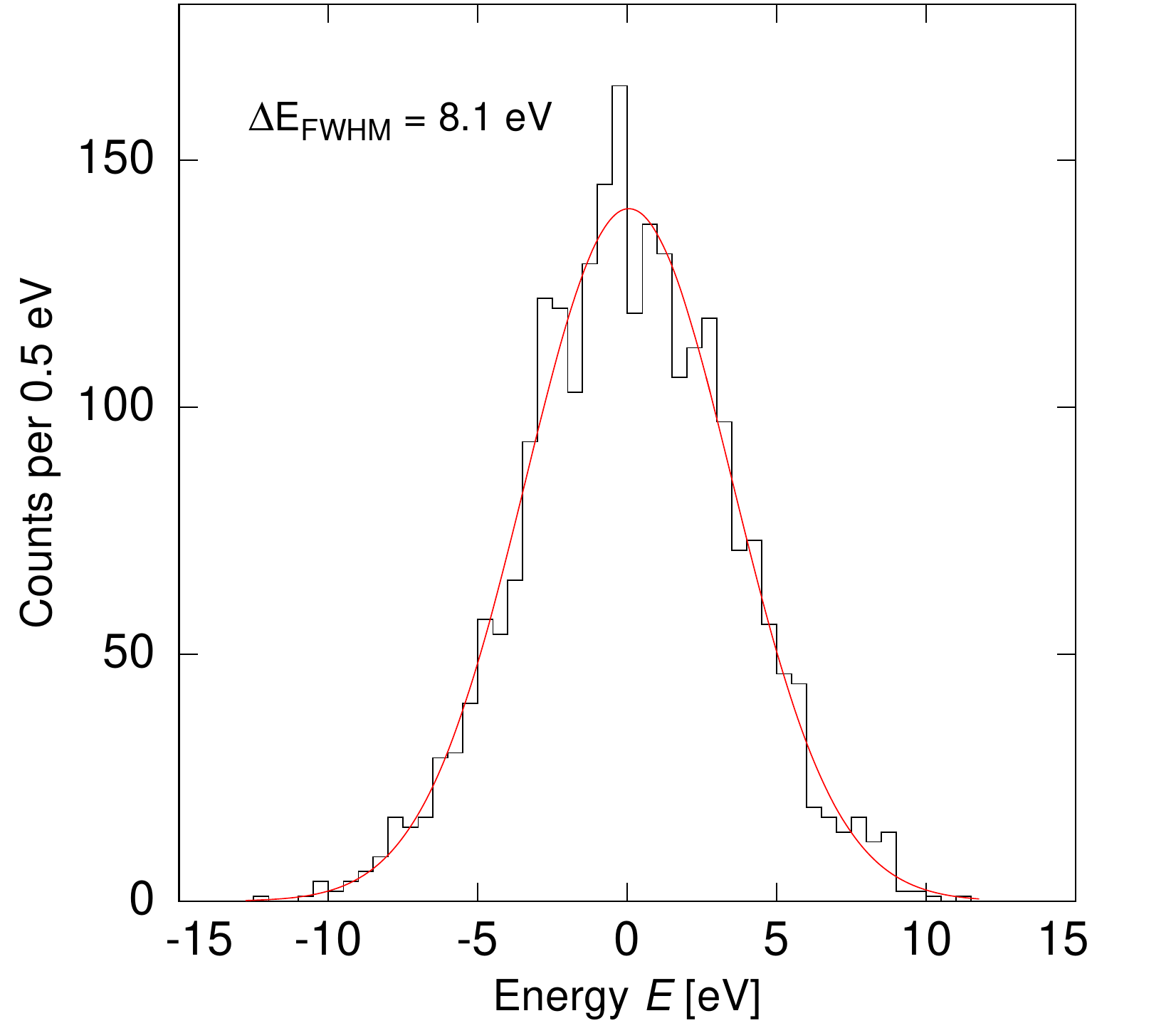}
  \caption{Baseline spectrum acquired with the MMC with ion-implanted absorber. A Gaussian fit with a width of 8.1 eV (FWHM) is superimposed to the experimental data (red line). \vspace{-1mm}}
  \label{baseline_Philipp}
\end{figure}
Fig. \ref{baseline_Philipp} shows the baseline spectrum of the
 detector with ion-implanted absorber operated at a temperature of $35\, \mathrm{mK}$ and with a
field generating current of $30\, \mathrm{mA}$.
The expected energy resolution for optimal operation conditions,
considering the design characteristics of the detector, including the extra heat capacity of $0.8 \,$pJ/K and typical
read-out performance, is $\Delta E_{\mathrm{FWHM}}\,=\,6.6\,$eV.
The fit of the baseline spectrum shows an energy resolution  of $\Delta E_{\mathrm{FWHM}}\,=\,8.1\,$eV.
The degradation of the energy resolution can be explained by the non-perfect experimental condictions that lead to additional noise contributions respect the one used in the theoretical calculations. A better design of the experiment and a better screening towards external source of noise will reduce the discrepancy.

\begin{figure*}
\centering
\includegraphics[width=0.48\textwidth]{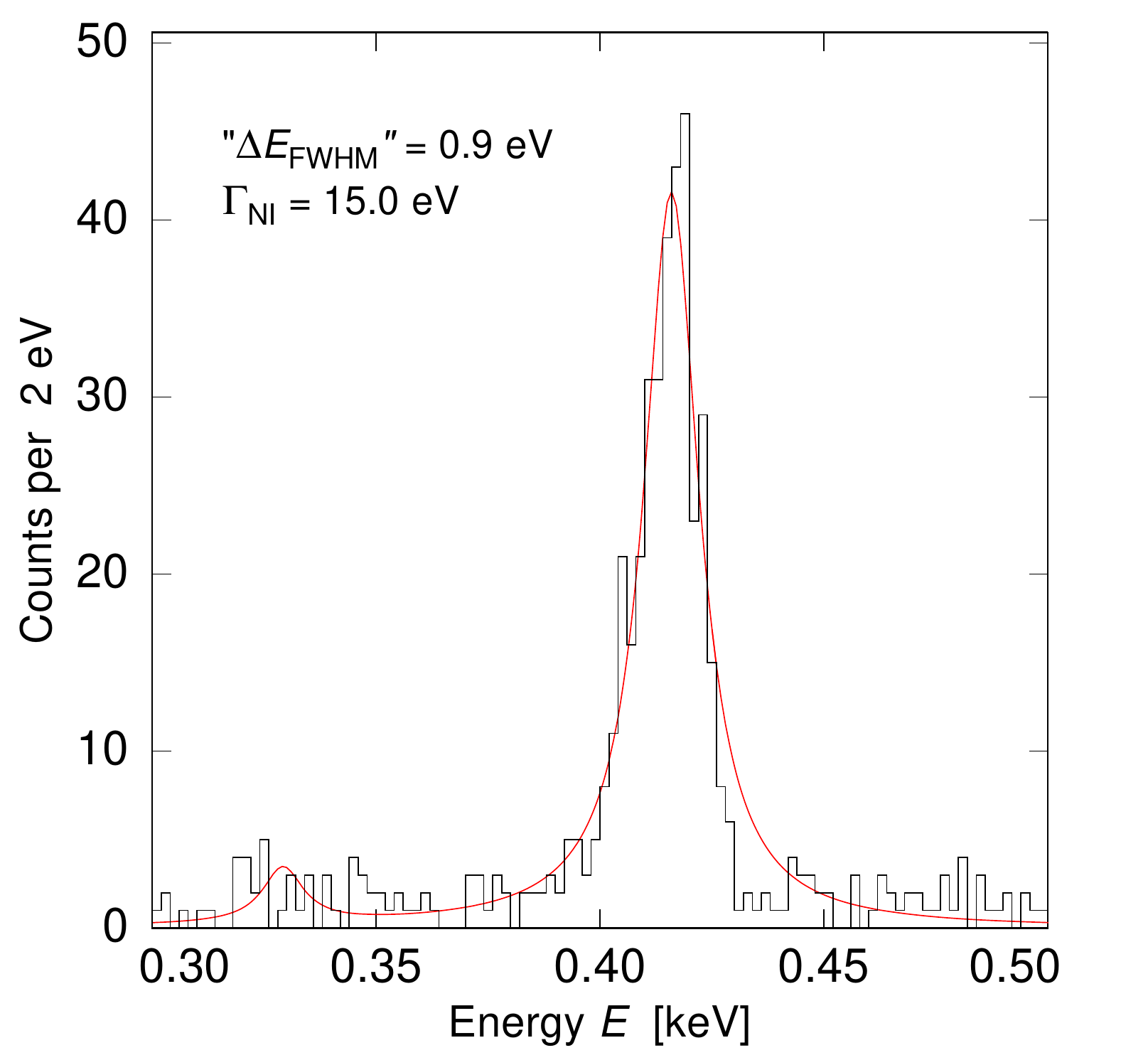}
\hfill
\includegraphics[width=0.48\textwidth]{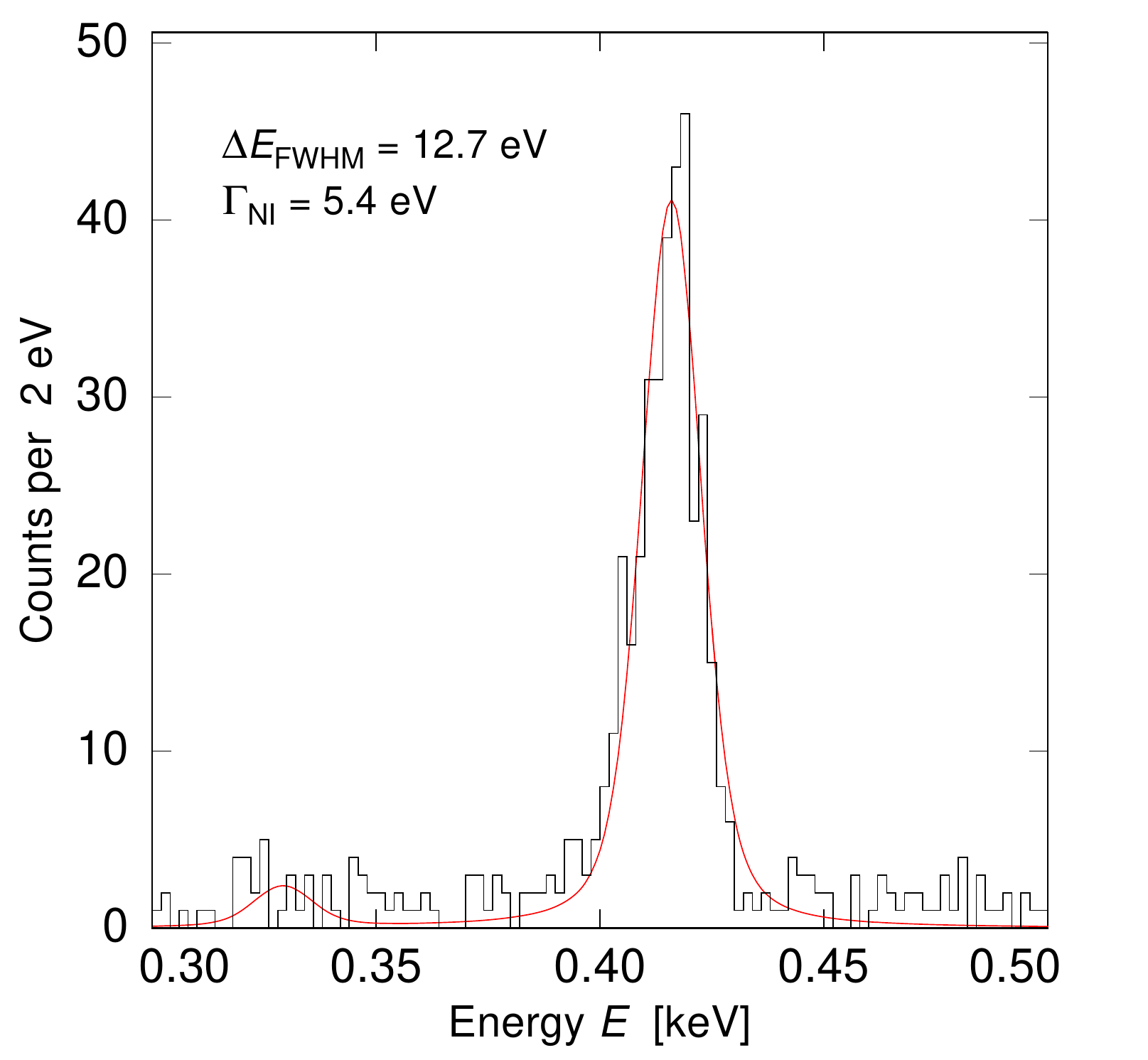}\\
\caption{Spectrum of the NI line of Dysprosium acquired with the MMC having an ion-implanted absorber. In a) the fit function, superimposed to the experimental data, is obtained by convolving the natural line-shape described by the parameters given in \citep{163Ho} with a gaussian having a width of 0.9 eV (FWHM), representing the detector response. In b) the fit function, superimposed to the experimental data, is obtained by convolving the natural line-shape described by the parameters given in \citep{Width_2001} with a gaussian having a width of 12.7 eV (FWHM).\vspace{-3mm}}
\label{NI_old}
\end{figure*}

In this experiment the rate of x-rays coming from
the external $^{55}$Fe calibration source was unintentionally kept very small, about one count
each 10 min. Due to the corresponding low statistics, the $K_{\alpha}$-line of manganese could not be used to define
the energy resolution at 5.9 keV, still it was used as a reference for the
energy calibration. To describe the resolution achievable in the energy range
of the measured spectrum, i.e. between $0\,$keV and $6\,$keV, the calorimetric NI line of dysprosium
was analyzed.
Fig. \ref{NI_old} a) and b) show the same experimental data for the calorimetric NI line of dysprosium but with two different fit functions superimposed. In both cases the fit based on the convolution of the natural line shape of the calorimetric NI line of dysprosium with a gaussian of variable width, representing the detector response, was used. In Fig. \ref{NI_old} a) the linewidth, $\Gamma_{\mathrm{NI}}\,=\,15\,$eV, suggested in \citep{163Ho},  was used to describe the natural shape of the line. It is clear that this fit shows neither a qualitatively satisfying agreement with the data nor a realistic detector response, since an energy resolution of $\Delta E_{\mathrm{FWHM}}\,=\,0.9\,$eV is impossible for this specific detector.
In Fig. \ref{NI_old} b) the theoretical line shape of the NI line of dysprosium was described using the linewidth, $\Gamma_{\mathrm{NI}}\,=\,5.4\,$eV, reported in \citep{Width_2001}.
The fit shown in Fig. \ref{NI_old} b) describes very well the measured data and the resulting energy resolution of $\Delta E_{\mathrm{FWHM}}\,=\,12.7\,$eV is in better agreement with the measured baseline resolution. The increased value for the gaussian width in the fit of the NI line of dysprosium compared to the gaussian width of the baseline spectrum fit could be explained by the relatively large temperature variations of the experimental holder during the experiment. The escape of high energy phonons is not so important for the NI line of neodymium due to the relatively low energy and due to the fact that part of the de-excitation energy is released through electron emission.
The comparison of Fig. \ref{NI_old} a) and b) gives a very important message, namely that the parameters describing the calorimetric measurement of the EC spectrum of $^{163}$Ho should still be more precisely investigated from experimental and theoretical side, in particular for the analysis of the EC spectrum in experiments devoted to the investigation of the neutrino mass. In this context, high precision calorimetric measurements can help testing the theoretical predictions.

From the analysis of the NI line we could estimate that the number of $^{163}$Ho ions that have been implanted in each absorber during this first implantation test performed at ISOLDE-CERN is about $2\times10^9$.

The comparison between the analysis of the energy resolution achieved by the detector with non-implanted absorber and the detector with ion-implanted absorber shows that no unexpected features of the treated detector were compromising the measured spectrum. This result follows also from the fact that the measured pulse height was not affected by any noticeable additional contribution to the specific heat.

In future detector developments, a new absorber concept will be studied.
The goal will be to reduce the heat capacity, in order to improve the intrinsic energy resolution of the detector, while keeping the quantum efficiency to be approximately 100$\%$. In order to achieve this, we will investigate new materials with reduced specific heat, like superconductors, to be used as absorber. A second important improvement will be to connect the absorber to the sensor through stems in order to eliminate the loss of energy by initial hot phonons.

\section{Energy calibration and linearity}
\begin{figure}[t]
  \includegraphics[angle=0, width=.44\textwidth]{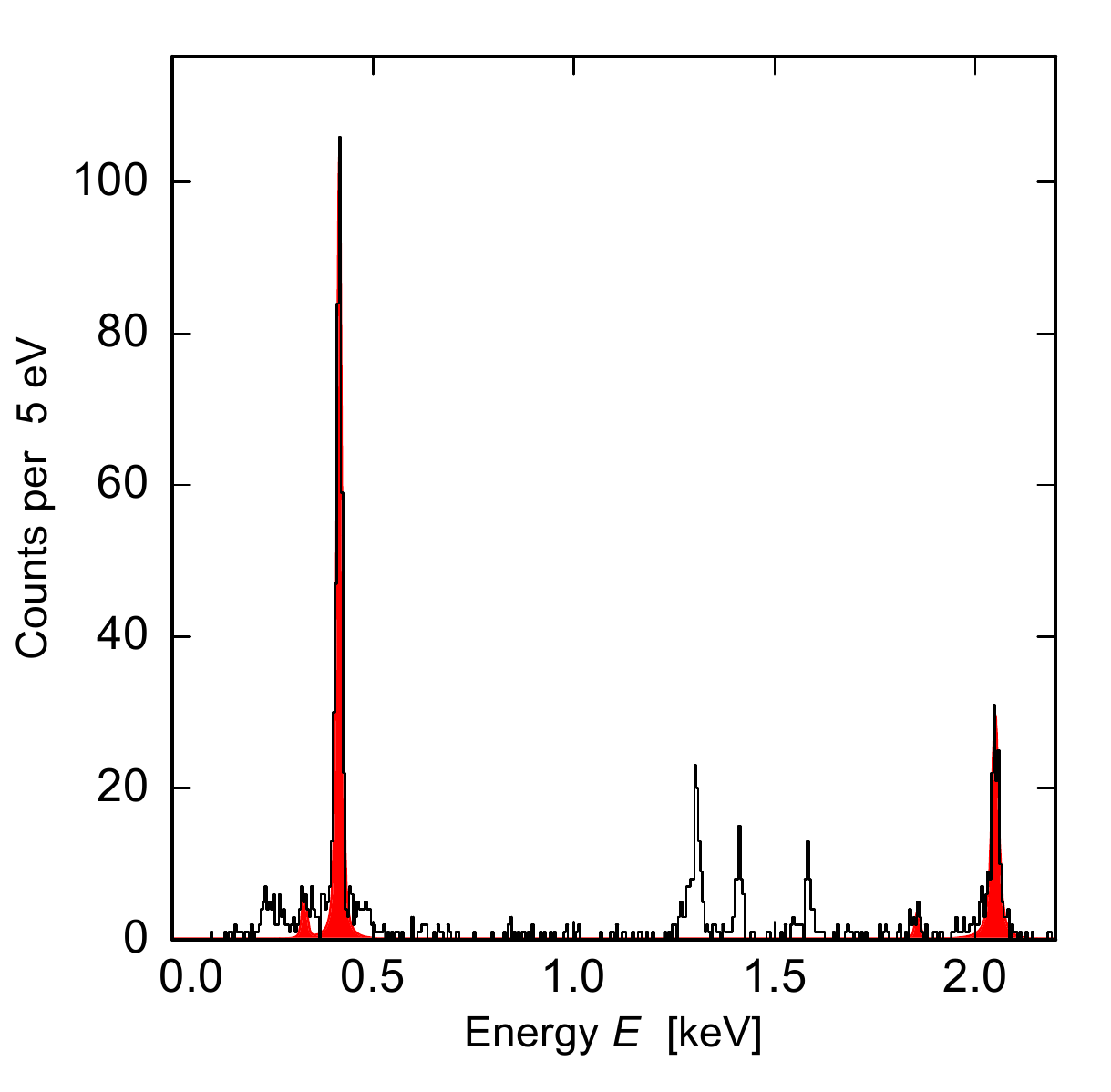}
  \caption{Calorimetric EC spectrum of $^{163}$Ho indicated by the superimposed fit (dashed region). The other lines which are present in the spectrum correspond to the calorimetric EC spectrum of $^{144}$Pm.
    \vspace{-5mm}}
  \label{Ho_spe}
\end{figure}

\begin{figure} [t]
  \includegraphics[angle=0, width=.48\textwidth]{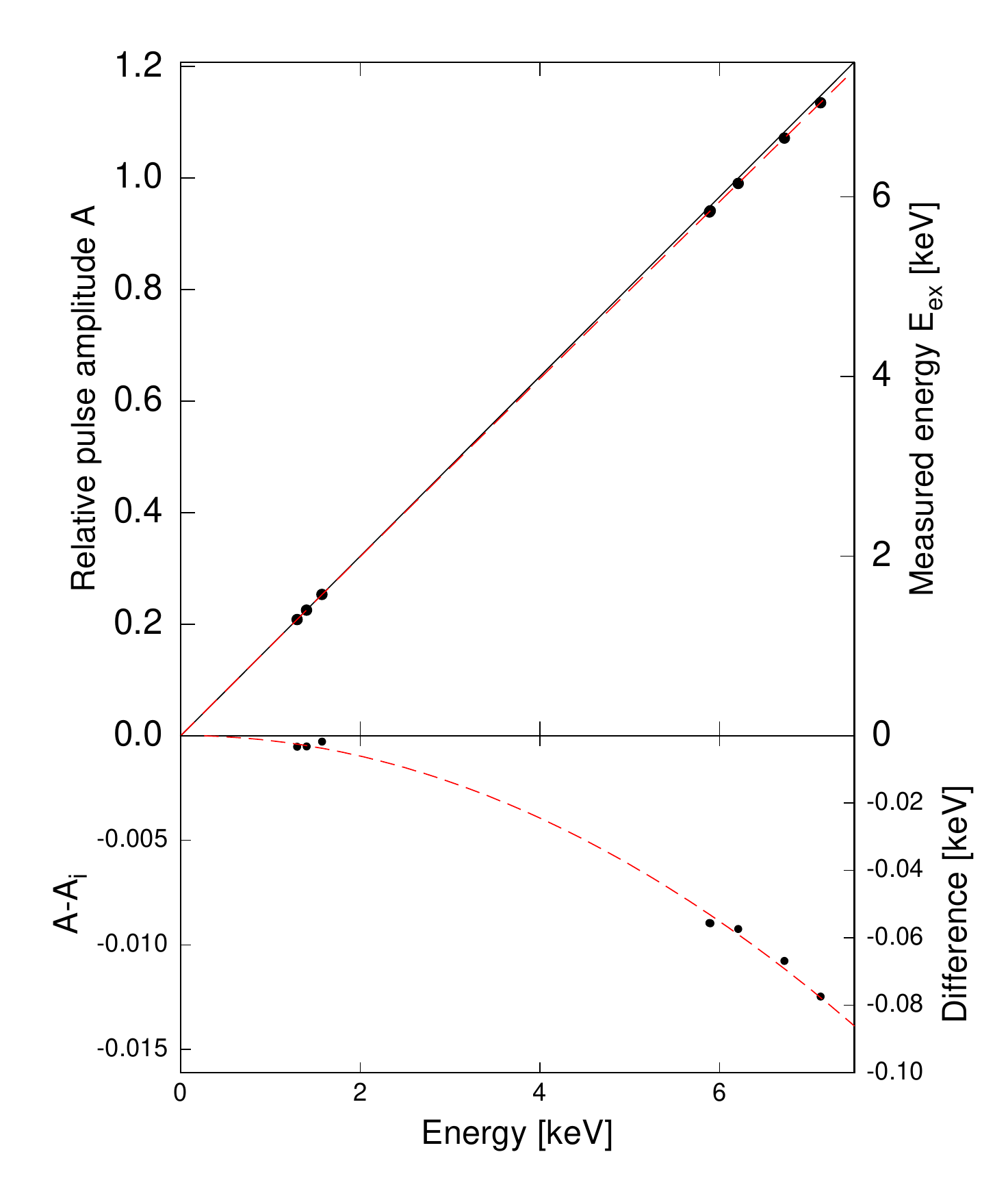}
  \caption{Energy calibration of the detector having ion-implanted absorber. In the upper part the symbols represent the measured signal amplitudes plotted versus the energies. A linear fit (continuous line) and a parabolic fit (dashed line) are superimposed to the data. In the lower part the detector the experimental amplitudes minus the corresponding values extracted from the linear fit are plotted versus energy with superimposed the quadratic part of the parabolic fit (dashed line). \vspace{-1mm}}
  \label{linearity}
\end{figure}

For the precise characterization of energy
spectra, the energy calibration and a confident understanding of the non-linearity of the detector are
extremely important. Fig. \ref{Ho_spe} shows the calorimetric measurement of the EC spectrum of $^{163}$Ho. The lines corresponding to the EC spectrum of $^{163}$Ho are marked by superimposing to the histogram the corresponding fit function represented by a shaded area The other lines, which have not superimposed fit function, represent the EC spectrum of $^{144}$Pm. More details for the description and analysis of the EC calorimetric spectrum of $^{163}$Ho can be found in \citep{Neutrino_LTD14}. Here, we want to discuss the detector linearity
in the energy range up to about $8\, $keV. The lines used to test the
linearity are the neodymium MI, MII, MII, LI, LII, LIII lines and the manganese
K$_{\alpha}$-line. In the upper part of Fig. \ref{linearity} the
measured signal amplitudes are plotted versus the energies reported in \citep{Thompson}. The symbols represent the experimental data. The continuous black line
represents a linear fit to the experimental points while the dashed
line represents a quadratic fit.

The lower part of Fig. \ref{linearity} shows the non-linearity of the
detector, as the difference between measured values and
values extracted by the linear part of the quadratic fit. As can be seen a non-linearity of less than 1$\%$ is
present at 8 keV. This value is in agreement with previous
measurements \citep{FleischLTD13}. With this energy calibration it
is possible to investigate the position of the calorimetric lines with the precision given by the energy resolution and the acquired statistics.

\section{Background from events in neighboring pixels and substrate}
As described in Section 2 all four absorbers were exposed to
the ion beam. Therefore in all the four absorbers radioactive decays are
occurring. Since the energy absorbed in a given pixel is then
transferred to the common silicon substrate, part of this heat can as well flow back to the other three detectors giving rise to background pulses. These spurious
thermal signals will appear in the energy spectrum of the read out detector, after being calibrated using the direct events, at a position given by the relative amplitude. These events can be well discriminated by the different pulse
shape and be can rejected. The largest contribution to the spurious events in case of the measured detector is due to the presence of the 40 keV atomic de-excitation following the capture of a K electron in $^{144}$Pm.
In Fig. \ref{4pixels} c) the letter A indicates the detector that was measured in this experiment. Within the thermal pulses that were acquired,
we could observe the direct events in the absorber of detector A, the spurious events due to heat flowing through the substrate after a K capture in the neighboring pixels and spurious events corresponding to interaction of those photons in the substrate due to the non-complete quantum efficiency of the absorber for the 40 keV radiation.

According
to the position of the detectors on the chip, we expected to see, in a plot
discriminating the pulse shape, four different families of pulses.
These four families represent the direct events in detector A, events in detector C (as indicated in Fig. \ref{4pixels} c))
positioned on the same diagonal as detector A, events in the two pixels, B and D, positioned on the other
diagonal, which have symmetric positions with respect to the detector A and can not be discriminated, and events in the substrate. Considering that, in first approximation, in each absorber the same number of ions has been implanted and using the quantum efficiency related to the absorber geometry and substrate volume, we quantified the number of expected events for the detection of the 40 keV photons in the different families. In particular, the number of events in the
read out detector A should be the same, within the statistical
errors, as the ones in the detector C on the same diagonal while the
family of events generated in the two pixels B and D on the other
diagonal should have twice the amount of events. The number of
events expected in the substrate, calculated according to the
stopping power in the different materials, is about the same as the one expected for the sum of the events in the pixels B and D.

\begin{figure}[t]
  \includegraphics[angle=0, width=.44\textwidth]{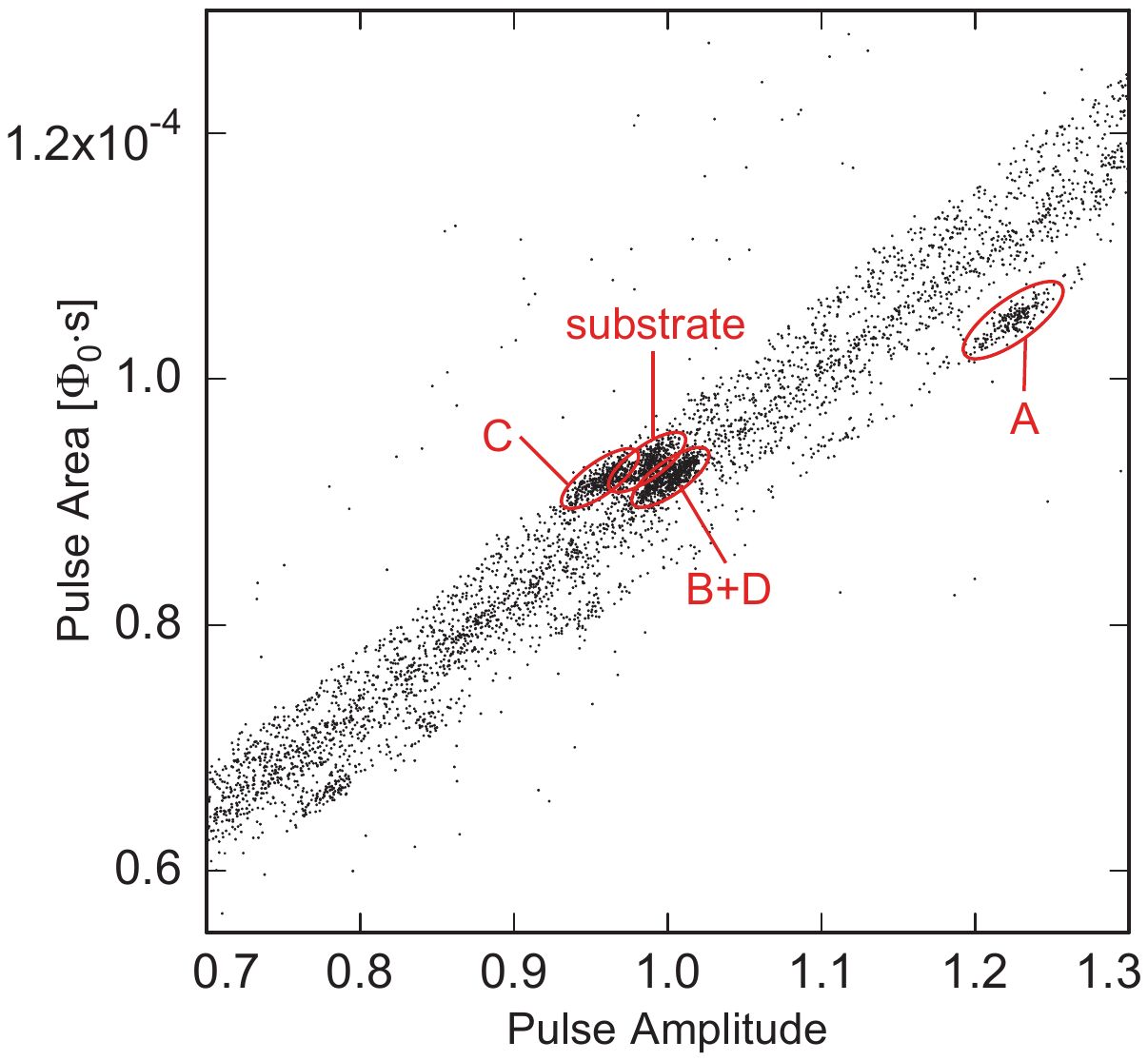}
  \caption{Plot of the area of the pulses, calculated as integral of the signal over the acquisition time-window versus
the fitted amplitude of the pulses with respect to an averaged reference pulse. Four families of pulses, the direct events in the read out detector (A), thermal signals from events in the neighboring pixels (C and B+D) and thermal signals from events in the substrate (Substrate) can be discriminated \vspace{-1mm}}
  \label{background}
\end{figure}

In Fig. \ref{background} the area of the pulses, calculated as integral of the signal over the acquisition time-window, is plotted versus
the fitted amplitude with respect to a reference pulse.
The four families of pulses can be well discriminated. For the direct events in detector A the group of points for the KI line of Nd is not contained in this amplitude scale (the group of events at amplitude 1.25 correspond to about 400 eV), while for the other three families the groups of points corresponding to KI are marked. As expected, since only part of the energy transferred to the substrate flows back to the read out detector, the amplitude of the spurious thermal pulses is much smaller than the thermal pulses generated by the same decay through direct events in the read out detector.
Analyzing the number of counts corresponding to what we
interpret as the Nd KI line, the expected ratios are obtained.
The number of events in the read out detector A is 370, in the
detector C in the same diagonal 365. For the two pixels B and D the number
of events is 740 and for the substrate 720.
This analysis shows that our model to discuss the different families of events in relation to their pulse shape agrees with the measured data. Moreover, the possibility to isolate families of events according to the location of the corresponding decay offers a useful method to reject the non-direct signals.
However, this method becomes less
efficient in the low energy range, below 100 eV (calibrated using the direct events in the read out detector), where the signal to
noise ratio is smaller. The difficulty in rejecting the
events occurring in the non-read out detector and in the substrate has the effect to increase the background which
complicates the analysis of the $^{163}$Ho EC spectrum.

In future detectors for a high precision calorimetric measurement of the EC spectrum of $^{163}$Ho, the presence of contaminants as $^{144}$Pm
will be eliminated and, with that, also the major contribution of
background events. Methods to obtain a very pure $^{163}$Ho source are already developed by our collaborators at the SAHA Institute for Nuclear Physics in Kolkata \citep{Susanta}.

Still to consider are the cross-talk pulses caused by the $^{163}$Ho decays occurring in the neighboring pixels. As an example, the MI transition at $2.046\, $keV for events occurring in the neighboring pixels,
is seen in the read out detector at a reduced signal amplitude corresponding to about 20 eV. This cross-talk
can degrade the energy resolution of the detector since it
could lead to non-resolvable pile-up. In addition natural
radioactivity and cosmic rays can produce events in the
substrate as well as in the detectors and therefore generate additional background.
A possible way to avoid background events generated through the substrate, such as cross-talk events from decays in neighboring pixels and the interaction in the substrate of particle originated from natural radioactivity,
would be the use of the full gradiometric meander pick-up coil structure \citep{PiesLTD14}. The cross-talk between neighboring pixels can be reduced by two orders of magnitude by having both meander-shaped pick-up coils equipped with sensor and absorber. A further improvement to reduce even more the problem of cross-talk would be the application of a membrane as detector substrate. In this case meander-shaped pick-up coils with the implanted detector on top will be positioned on
a SiN membrane which is only about $1\,\mu$m thick. Suitable thermal links, made out of thin normal metal films, will connect the detector to the thermal
bath. All the pixels will then be thermally disconnected from each
other and from the Si substrate.

\section{Conclusion}

In this work we have characterized a fully micro-fabricated MMC
which was irradiated at ISOLDE. We compared the performance of the detector with the implanted absorber with the performance of a detector with non-implanted absorber. We could demonstrate that the performance of the detector, after undergoing the ion-implantation process
is not degraded. Moreover, we
showed the possibility to measure with high energy resolution
the atomic de-excitation following an EC decay of implanted ions, $^{163}$Ho and $^{144}$Pm.
In particular we described the case of the NI line of dysprosium. We discussed the
importance of high resolution calorimetric measurements to investigate the line shape of the atomic de-excitation analyzing the case of the NI
line of dysprosium.

The performance achieved by the tested detector proves that
MMCs with the absorber having ion-implanted $^{163}$Ho source, can provide high energy resolution calorimetric measurements of the
EC spectrum of $^{163}$Ho. We pointed out possible improvements in the detector design which will lead to an improved energy sensitivity and to a reduction of background events. The optimization of future detector designs will be a fundamental aspect of the ECHO experiment \citep{ECHO} aiming at the investigation of the electron neutrino mass in the sub-eV energy range.

\vspace{5mm}

This work was supported by the FRONTIER program of Heidelberg
University.








\end{document}